\documentclass[prb,twocolumn,amsmath,amssymb,floatfix,superscriptaddress,aps]{revtex4-2}

\usepackage{color}
\usepackage{graphicx}
\usepackage{hyperref}
\usepackage{braket}
\usepackage[capitalize]{cleveref}
\usepackage[normalem]{ulem}
\usepackage{bbm}
\usepackage{soul}

\newcommand{\Tr}{{\rm Tr}}
\renewcommand{\Im}{{\rm Im}}
\newcommand{\pp}{{\prime\prime}}

\begin{document}

\title{Fully analytical equation of motion approach for the double quantum dot in the Coulomb blockade regime}

\author{Nahual Sobrino}\email{nahualcarlos.sobrinoc@ehu.eus}
\affiliation{Nano-Bio Spectroscopy Group and European Theoretical Spectroscopy Facility (ETSF), Departamento de Pol\'imeros y Materiales Avanzados: F\'isica, Qu\'imica y Tecnolog\'ia, Universidad del Pa\'is Vasco UPV/EHU, Avenida de Tolosa 72, E-20018 San Sebasti\'an, Spain}

\author{David Jacob}
\affiliation{Nano-Bio Spectroscopy Group and European Theoretical Spectroscopy Facility (ETSF), Departamento de Pol\'imeros y Materiales Avanzados: F\'isica, Qu\'imica y Tecnolog\'ia, Universidad del Pa\'is Vasco UPV/EHU, Avenida de Tolosa 72, E-20018 San Sebasti\'an, Spain}
\affiliation{IKERBASQUE, Basque Foundation for Science, Plaza Euskadi 5, E-48009 Bilbao, Spain}

\author{Stefan Kurth}
\affiliation{Nano-Bio Spectroscopy Group and European Theoretical Spectroscopy Facility (ETSF), Departamento de Pol\'imeros y Materiales Avanzados: F\'isica, Qu\'imica y Tecnolog\'ia, Universidad del Pa\'is Vasco UPV/EHU, Avenida de Tolosa 72, E-20018 San Sebasti\'an, Spain}
\affiliation{IKERBASQUE, Basque Foundation for Science, Plaza Euskadi 5, E-48009 Bilbao, Spain}
\affiliation{Donostia International Physics Center (DIPC), Paseo Manuel de
	Lardizabal 4, E-20018 San Sebasti\'{a}n, Spain}
\date{\today}

\begin{abstract}
  A fully analytical approach based on the equation of motion technique to
  investigate the spectral properties and orbital occupations in an
  interacting double quantum dot in equilibrium is presented.
  By solving a linear system for the density correlators
  analytically, an explicit expression for the one
  body Green's function in terms of local occupations, intra- and inter-dot
  Coulomb interactions, and the model parameters is derived. In the
  uncontacted limit, the results coincide with those obtained from the grand canonical ensemble.
  The analytical results compare favorably with
  numerical results obtained with the non-crossing approximation and the
  hierarchical equation of motion methods accurately reproducing
  peak positions and spectral weight distributions in the Coulomb blockade regime.
\end{abstract}

\maketitle
 
\section{Introduction}

In the last decades quantum dots (QDs) have attracted  an enormous amount of attention due to their potential applications in quantum computing, nanoelectronics, and as ideal systems to
explore fundamental quantum phenomena\cite{hanson2007, kloeffel2013, loss1998, michler2000, petta2005, reimann2002, utic2004, veldhorst2015, zwanenburg2013}.
QDs are essentially nano\-structures that confine electrons within a small region of space, leading
to the quantization of the electronic motion in all dimensions. They can be realized in manifold ways, e.g.,
by electrostatic confinement of a two-dimensional electron gas~\cite{hanson2007}, metallic or semiconductor
nanoparticles~\cite{petta2001studies,pelayo2021semiconductor}, carbon nanotubes~\cite{nygard2000kondo,buitelaar2002multiwall},
or the buckminster fullerene (C$_{60}$)~\cite{parks2007tuning}.

 A QD coupled to electrodes can be effectively described by an Anderson impurity model (AIM)\cite{anderson1961localized}, which has been a cornerstone for understanding strongly correlated
electronic systems. The AIM offers an ideal playground to explore electron-electron interactions, shedding light on phenomena
such as the Kondo effect and magnetic impurity behavior in metals, significantly advancing our understanding of many-body
physics\cite{hewson1997}. These insights have paved the way for exploring more complex systems, such as double quantum dots (DQDs),
which offer a unique platform for investigating electron-electron interactions, quantum coherence, and transport properties in
low-dimensional systems. The coupling between QDs  and the presence of interdot interactions introduce complex and
intriguing physics that have been extensively studied in and out of equilibrium both theoretically and experimentally
\cite{ono2002current,vanderwiel2002electron,you1999electron,zhang2023inverse,petta2004manipulation,chen2004transition,fernandez2022quantum,bouman2020triplet}.

  Various computational approaches have been developed for solving the AIM and its extensions, among them
numerical renormalization group (NRG), quantum Monte Carlo (QMC), non-crossing approximation (NCA),
and hierarchical equations of motion (HEOM)\cite{bulla2008,gull2011,aoki2014nonequilibrium,schollwock2011,coleman1984nca,bickers1987,eckstein2010,tanimura2006stochastic,ishizaki2009unified}.
 On the other hand, the equation of motion (EOM) approach
has emerged as a powerful tool for investigating Green's functions in  Hubbard and
  Anderson impurity models\cite{zubarev1960double,hubbard1963electron,hubbard1964electronII,hubbard1964electronIII,meir1992}.
The EOM approach provides a systematic way to derive the equations governing the dynamics of the Green's
functions (GFs), allowing for a detailed analysis of the spectral properties and electronic correlations.
This approach has been successfully applied to study the spectral and transport properties of single and
multiple quantum dot systems under various interaction regimes\cite{van2010anderson, swirkowicz2003nonequilibrium, you1999spectral, sztenkiel2007electron, kuo2007tunneling, lamba2000transport, palacios1997fine, sierra2016interactions, bulka2004electronic, sun2002double, chi2006interdot,chang2008theory,sierra2016interactions,feng2011fast}. However, the
complexity of the equations due to inter-dot interactions and coupling to
electronic reservoirs required different approximations
and numerical techniques  to accurately describe
correlated electronic states and transport phenomena. 

Despite the progress made using numerical techniques, an analytical approach
offers the distinct advantage of providing explicit functional dependencies.
Understanding these dependencies is crucial for designing and optimizing quantum
dot-based devices, particularly in regimes where strong electron-electron interactions
play a dominant role. Furthermore, analytical solutions offer a computationally efficient
 way to study additional properties and explore new physical regimes,
guiding experimental efforts and advancing the development of quantum technologies.

The rest of the paper is organized as follows: In Sec.~\ref{model_eom}, we first introduce the
DQD Hamiltonian coupled to reservoirs, and explain in detail our analytic solution for the DQD via the EOM approach.
In Sec.~\ref{results}, we provide a comprehensive comparison of the analytical results with numerical methods,
discussing the accuracy and reliability of the analytical EOM approach.
The conclusions are presented in Sec.~\ref{conclus}. Finally, we provide details of the derivations
in three Appendixes: Appendix~\ref{app_derivation_approx} details our
approximate treatment of the leads, Appendix~\ref{app_Green_functions}
gives details for the different GFs while Appendix~\ref{app_correlators}
presents the linear system for the density correlators and its analytical solution.

\section{Model and Method}
\label{model_eom}

\subsection{Double quantum dot model}

We consider a parallel double quantum dot (DQD) system, where each quantum dot
(QD) $i=1,2$ is attached to its own left (L) and right (R)
 reservoirs of non-interacting
electrons. The Hamiltonian of the system has the form

\begin{align}
  \hat{\mathcal{H}} &= \sum_i v_{i}\, \hat n_{i} + \sum_i U_{i} \, \hat n_{i\sigma}\hat n_{i\bar\sigma}+ U_{12} \, \hat n_{1}\hat n_{2} \nonumber\\
  &+\sum_{\alpha k \sigma}\epsilon_{\alpha k  i }
    \,\hat c^{\dagger}_{\alpha  k i \sigma}\hat c_{\alpha  k i\sigma}
    +\sum_{i \alpha k \sigma } \left( V_{\alpha k i} \, \hat c^{\dagger}_{ \alpha k i \sigma}\hat d_{i \sigma}+\text{H.c.}\right)\;,
    \label{eq_H}
\end{align}
where $\hat d_{i\sigma}$ ($\hat d_{i\sigma}^\dagger$) is the annihilation (creation) operator for an electron
with spin $\sigma$ on QD $i$, and $\hat c_{\alpha k i\sigma}$ ($\hat c_{\alpha k i \sigma}^\dagger$) is the annihilation
(creation) operator for an electron in state $k,\sigma$ in reservoir $\alpha=\text{L,R}$ connected to dot $i$.
$\hat n_{i}=\sum_\sigma\hat n_{i\sigma}$ and $\hat n_{i\sigma}=\hat d^{\dagger}_{i\sigma}\hat d_{i\sigma}$
are the total and the spin resolved occupation operators for QD $i$, respectively.

The first three terms in \cref{eq_H} describe the DQD with on-site energies (or gates) $v_i$
and subject to intra- and inter-dot Coulomb repulsion $U_{i}$ and $U_{12}$, respectively.
The last two terms in \cref{eq_H} describe the single-particle eigenstates of the
reservoirs with energies $\epsilon_{\alpha k i  }$
and the coupling to the corresponding QD with hopping $V_{\alpha ki}$.


\subsection{Analytical equation of motion approach}

We now apply the EOM approach to obtain an analytic expression for the single-particle GF solely in terms of
the occupations $\langle\hat{n}_i\rangle$ of each QD $i$ and the electron addition and removal energies. The derived expression
is valid for the weak-coupling regime, i.e. when the coupling to the reservoirs is small compared to the temperature, and
becomes exact in the limit of vanishing coupling, in contrast to other approaches, which usually rely on some kind of mean-field
decoupling~\cite{palacios1997fine,sierra2016interactions}.
Our approach consists of two steps: First the EOMs are solved recursively,
leading to expressions for both the single-particle as well as higher-order
GFs in terms of density correlators
$\langle \hat{n}_{i_1\sigma_1} \hat{n}_{i_2\sigma_2} \ldots\rangle$ 
where the brackets $\braket{\dots}$ indicate the thermal average, i.e.,
$\braket{\hat{\mathcal{A}}}=\Tr[e^{-\beta\hat{\mathcal{H}}}\hat{\mathcal{A}}]/\Tr[e^{-\beta\hat{\mathcal{H}}}]$
with the inverse temperature $\beta\equiv 1/T$.  
From the equations for the GFs, in a second step we obtain a set of linear equations for the correlators, which in the case of the DQD can be solved
analytically. The solution allows us to express the correlators and thus the GF in terms of the dot occupations
$\langle\hat{n}_i\rangle$.

The retarded single-particle GF of the DQD system given in \cref{eq_H} is defined in the time domain as
\begin{align}
  \label{eq_G_t}
  G_{i\sigma}^{r}(t) = - i \theta(t) \braket{ \{\hat d_{i \sigma}(t), \hat d_{i\sigma}^{\dagger}(0)\}}\;,
\end{align}
where $\theta(t)$ is the Heaviside step function, the curly
  brackets denote the anticommutator, and for any operator $\cal{\hat{A}}$ 
  the corresponding operator in the Heisenberg picture is 
  $\hat{\cal{A}}(t) =
  \exp(i {\cal{\hat{H}}}t) \hat{\cal{A}} \exp(-i {\cal{\hat{H}}}t)$.
The Fourier transform of \cref{eq_G_t}   
\begin{align}
  G_{i\sigma}^{r}(\omega) \equiv\braket{\braket{\hat d_{i\sigma}:\hat d^{\dagger}_{i\sigma}}}_{\omega} = \int_{-\infty}^{\infty} G_{i\sigma}^{r}(t) \, e^{i\omega t}dt
  \label{eq_G_w}
\end{align}
can be directly related to the orbital occupation via
($\int \equiv \int\frac{d\omega}{2\pi}$ in the following)
\begin{align} 
  \braket{\hat n_{i\sigma}} = -\int f(\omega) \, \text{Im}\left(G_{i\sigma}^{r}(\omega)\right)\;,
  \label{eq_exp_values1}
\end{align}
where $f(\omega)= [1+e^{\beta\omega}]^{-1}$ is the Fermi function.

In order to proceed, it is convenient to introduce a more general GF that represents the GF generated at all orders of the EOM approach:
\begin{align}
  G_{B}^{r}(t) = - i \theta(t) \braket{ \{\hat{B}(t), \hat d_{i\sigma}^{\dagger}(0)\}}\;,
  \label{eq_G_general_t}
\end{align}
where the operator $\hat B(t)$  is a product of occupation operators multiplied by either a single dot annihilation operator $\hat{d}_{i\sigma}$
or a lead annihilation operator $\hat{c}_{\alpha ki\sigma}$ in the Heisenberg picture.  
The EOM of its Fourier transform $G_{B}^{r}(\omega)\equiv\braket{\braket{\hat B:\hat d^{\dagger}_{i\sigma}}}_{ \omega}$ in the frequency domain can be written as~\cite{zubarev1960double,hewson1966theory,hubbard1964electronII}
\begin{gather}
  \omega^+ \braket{\braket{\hat B:\hat d^{\dagger}_{i\sigma}}}_\omega
  = \braket{\{\hat B,\hat d^{\dagger}_{i\sigma}\}} + \braket{\braket{[\hat B,\hat{\mathcal{H}}]:\hat d^{\dagger}_{i\sigma}}}_\omega\;,
  \label{eq_EOM_B}
\end{gather}
where $\omega^+\equiv\omega+i\eta$ with $\eta\rightarrow0^+$ is the energy
$\omega$ shifted infinitesimally to the complex plane.
We define the general $N+1$ particle dot GF as
$G_{D d_{i\sigma}}^{r}(\omega)\equiv\braket{\braket{\hat D\hat d_{i\sigma}:\hat d^{\dagger}_{i\sigma}}}_{ \omega}$ 
where $\hat{D}=\hat{n}_{i_1\sigma_1}\hat{n}_{i_2\sigma_2}\ldots\hat{n}_{i_N\sigma_N}$
with $(i_k,\sigma_k)\neq(i_q,\sigma_q)$ for all $k\neq q$ and
$(i_k,\sigma_k)\neq(i \sigma)$ for all $k\in \{1,\ldots,N\}$
and $\hat{D}=\hat{\mathbbm{1}}$ is the unit operator if $N=0$.

In the following we use the EOM to derive analytical
expressions for the dot equilibrium GFs $G^r_{D d_{i \sigma}}(\omega)$
of the DQD system attached to two  reservoirs solely in terms of the local
occupations
$\braket{\hat{n}_{i\sigma}}$ and the removal and addition energies of the DQD. 
Our starting point is the EOM for the retarded single-particle GF 
\begin{align}
  \omega^+ \, \braket{\braket{\hat d_{i\sigma}:\hat d^{\dagger}_{i\sigma}}}_\omega =&
  \braket{\{\hat d_{i\sigma},\hat d^{\dagger}_{i\sigma}\}} +\braket{\braket{[\hat d_{i\sigma},\hat{\mathcal{H}}]:\hat d^{\dagger}_{i\sigma}}}_\omega\;.
  \label{eq_EOM_GF}
\end{align}
The first term of the right-hand side (r.h.s.) of \cref{eq_EOM_GF} directly follows from the anti-commutation rules for
creation and annihilation operators: $\braket{\{\hat d_{i\sigma},\hat d^{\dagger}_{i\sigma}\}}=1$.
The commutator in the second term of the r.h.s. of \cref{eq_EOM_GF}  is found to be
\begin{gather}
  [\hat d_{i\sigma},\hat{\mathcal{H}}]=\left(v_i+U_{i}\,\hat n_{i\bar \sigma}+U_{12}\,\hat n_{\bar i}\right)\hat d_{i\sigma}+\sum_{\alpha k}V^{*}_{\alpha k i}\,\hat c_{\alpha k i\sigma}\;,
  \label{eq_Heis}
\end{gather}
where $\bar i$ refers to the site index opposite to $i$.
 Hence the second term of the r.h.s. of \cref{eq_EOM_GF} generates a contribution to the single-particle GF,  two higher order GFs and a mixed dot-lead GF according to:
\begin{gather}
  (\omega^+-v_i) \, G_{i\sigma}^{r}(\omega) = 1+U_{i}\braket{\braket{\hat n_{i\bar\sigma}\,\hat d_{i\sigma}:\hat d^{\dagger}_{i\sigma}}}_\omega \nonumber\\
  +U_{12}\braket{\braket{\hat n_{\bar i}\,\hat d_{i\sigma}:\hat d^{\dagger}_{i\sigma}}}_\omega+\sum_{\alpha k}V^{*}_{\alpha k i}\braket{\braket{\hat c_{\alpha k i \sigma}:\hat d^{\dagger}_{i\sigma}}}_\omega\;.
  \label{eq_EOM_1}
\end{gather}
The EOM of the mixed dot-lead GF is obtained from \cref{eq_EOM_B} with $\hat{B}=\hat c_{\alpha k i \sigma}$
\begin{align}
	\omega^+ \braket{\braket{\hat c_{\alpha k i \sigma}:\hat d^{\dagger}_{i\sigma}}}_\omega = \braket{\braket{[\hat c_{\alpha k i \sigma},\hat{\mathcal{H}}]:\hat d^{\dagger}_{i\sigma}}}_\omega\;,
\end{align}
where the commutator between the annihilation operator of the leads and the
Hamiltonian is
$[\hat{c}_{\alpha k i \sigma},\hat{\mathcal{H}}]=\epsilon_{\alpha k i}
\hat c_{\alpha k i \sigma}+ V_{\alpha k i}\hat d_{i \sigma}$. Then, the last term of the r.h.s of \cref{eq_EOM_1}  becomes
\begin{equation}
\sum_{\alpha k}V^{*}_{\alpha k i}\braket{\braket{\hat c_{\alpha k i \sigma}:\hat d^{\dagger}_{i\sigma}}}_\omega= \Delta_{i}(\omega) \braket{\braket{\hat d_{i\sigma} :\hat d^{\dagger}_{i\sigma}}}_\omega\;,
\label{eq_EOM_2}
\end{equation}
where we have defined the embedding self energy or hybridization function as
\begin{equation}
  \Delta_{i}(\omega) \equiv \sum_{\alpha k}
  \frac{|V_{\alpha k i}|^2}{\omega^+- \epsilon_{\alpha k i}} \;.
  \label{eq_Delta}
\end{equation}
The hybridization function $\Delta_i(\omega)$ describes the effect of the coupling of QD $i$ to its reservoirs.
More specifically, its real part yields the renormalization of the on-site energy $v_i$,
while its imaginary part describes the life-time broadening of the QD level due to the
coupling to the reservoirs. Note that due to the fact that 
	each dot only couples to its own left and right leads the embedding 
	self energy is diagonal in the dot indices.

Taking into account the definition of the hybridization function \cref{eq_Delta}, the single-particle GF can be written as 
\begin{gather}
	(\omega-v_i-\Delta_{i}(\omega) ) \, G_{i\sigma}^{r}(\omega) = 1+U_{i}\braket{\braket{\hat n_{i\bar\sigma}\hat d_{i\sigma}:\hat d^{\dagger}_{i\sigma}}}_\omega \nonumber\\
	+U_{12}\braket{\braket{\hat n_{\bar i}\hat d_{i\sigma}:\hat d^{\dagger}_{i\sigma}}}_\omega\;.
	\label{eq_EOM_3}
\end{gather}

 order GFs thus increases by one.

Setting up the EOMs for the higher order GFs on the r.h.s. of \cref{eq_EOM_3} leads, on the one hand,
to the generation of GFs of yet higher order of the form
$\braket{\braket{\hat n_{i^\pp\sigma^\pp} \hat n_{i^\prime\sigma^\prime} \hat d_{i\sigma}:\hat d^{\dagger}_{i\sigma}}}_\omega$. In each
subsequent order the number of occupation operators in the generated higher-order GFs thus increases by one.
On the other hand, it also generates mixed dot-lead GFs of the same
  order. Obviously, we somehow need to truncate the hierarchy of the EOMs.

  Below we introduce a truncation scheme, which can be justified in three
  different ways, all leading to the same result. We first consider the
  analytic structure of the (dot) GFs of the uncontacted system, which can be
  written as sums over single poles at certain frequencies shifted
  infinitesimally from the real axis. The residues of these poles are
  given by linear combinations of density correlators. A physically plausible
  model of approximately taking into account the coupling to the leads
  (and the first way of introducing our truncation scheme) is then given by
  simply broadening the poles by the embedding self-energies~\cite{ryndyk2009green,haug2008quantum,niu1999equation}. As a second
  way, the same approximation can also be obtained by neglecting certain GFs,
  which contain one or more lead operators (for a detailed derivation see
  Appendix \ref{app_derivation_approx}). Finally, the very same approximate
  truncation of the EOMs can also be achieved by assuming that the local
  density operator and the Hamiltonian commute, i.e.,
  $[ \hat{n}_{i\sigma},\hat{\mathcal{H}}]\approx0$, which becomes exact in the uncontacted situation.
The GFs calculated within this approximation are expected to be
accurate in the Coulomb blockade (CB) regime, when the temperature $kT$
is large compared to the broadening $-\Im\Delta_i$ by the reservoirs. 

Under this approximation, the mixed higher-order GFs are related to the same-order dot GFs via
\begin{align}
  \braket{\braket{\hat{D} \hat c_{\alpha k i \sigma}:\hat d^{\dagger}_{i\sigma}}}_\omega &\approx  \braket{\braket{\hat{D}[ \hat c_{\alpha k i \sigma},\hat{\mathcal{H}}]:\hat d^{\dagger}_{i\sigma}}}_\omega\nonumber\\
  &=\frac{V_{\alpha k i}}{\omega^+-\epsilon_{\alpha k i}}\braket{\braket{\hat{D}
      \hat d_{i\sigma}:\hat d^{\dagger}_{i\sigma}}}_\omega \;.
\end{align}
Therefore, the one-body dot GF \cref{eq_G_w} can be expressed
completely in terms of higher-order dot GFs $G_{Dd_{i\sigma}}^r(\omega)$. Hence,
for a finite number of single-particle levels the EOM can be closed at finite order. The EOM approach can thus be solved by recursive evaluation of the EOMs for successive orders of GFs. 
We can easily see that for the DQD the EOMs can be closed at the fourth order $G_{D d_{i\sigma}}^{r}(\omega)$ where $\hat{D}=\hat{n}_{i\bar\sigma}\hat{n}_{\bar{i}\sigma}\hat{n}_{\bar{i}\bar{\sigma}}$,
since the additional operator $\hat{n}_{i^\prime\sigma^\prime}$  generated by the interaction part of the EOM must be equal to either one of
the operators $\hat{n}_{i\bar\sigma}$, $\hat{n}_{\bar{i}\sigma}$ or $\hat{n}_{\bar{i}\bar\sigma}$ in $\hat{D}$. 
This yields the four-particle GF
\begin{equation}
  \braket{\braket{\hat n_{i\bar\sigma}\hat n_{\bar i\sigma}\hat n_{\bar i\bar \sigma}\hat d_{i\sigma}:\hat d^{\dagger}_{i\sigma}}}_\omega
  =\frac{\braket{\hat n_{i\bar\sigma}\hat n_{\bar i\bar\sigma}\hat n_{\bar i\sigma}}}{\omega-v_i-U_i-2 U_{12}-\Delta_i(\omega)}\;.
  \label{4part_gf}
\end{equation}

As shown in Appendix~\ref{app_Green_functions},
replacing the expressions of the higher-order GFs (expressed in terms of the correlators \cref{eq_exp_values})
recursively into the lower-order GFs finally allows us to express the one-body GF (\ref{eq_G_w}) as 
\begin{align}
  &G_{i\sigma}^r(\omega)=\sum_{j=1}^{6}\frac{r_{i,j}}{\omega-p_{i,j}-\Delta_i(\omega)}\;,
    \label{eq_GF0}
\end{align}
where in order to simplify the notation, we have introduced the following abbreviations for the
addition and removal energies, which define the poles of the GF (\ref{eq_GF0}):
\begin{subequations}
	\begin{align}
		&p_{i,1}=v_{i}\;,\\
		&p_{i,2}=v_{i}+U_{i}\;,\\
		&p_{i,3}=v_{i}+U_{i}+U_{12}\;,\\
		&p_{i,4}=v_{i}+U_{i}+2\,U_{12}\;,\\
		&p_{i,5}=v_{i}+U_{12}\;,\\
		&p_{i,6}=v_{i}+2\,U_{12}\;.
	\end{align}
	\label{eq_pi}
\end{subequations}

Importantly, the residues $r_{i,j}$ in \cref{eq_GF0} can be expressed by linear combinations of the one-, two-,
and three-body correlators as follows:
\begin{subequations}
  \begin{eqnarray}
    r_{i,1} &=& 1-\braket{\hat n_{i\sigma}}-2\braket{\hat n_{\bar i\sigma}}+\braket{\hat n_{\bar i\sigma}\hat n_{\bar i\bar\sigma}}
              +2\braket{\hat n_{i\bar\sigma}\hat n_{\bar i\bar\sigma}}\nonumber\\
          && -\braket{\hat n_{i\bar\sigma}\hat n_{\bar i\bar\sigma}\hat n_{\bar i\sigma}}\;, \\
    r_{i,2} &=& \braket{\hat n_{i\sigma}}-2\braket{\hat n_{i\bar\sigma}\hat n_{\bar i\bar\sigma}}+\braket{\hat n_{i\bar\sigma}\hat n_{\bar i\bar\sigma}\hat n_{\bar i\sigma}}\;,\\
    r_{i,3} &=& 2(\braket{\hat n_{i\bar\sigma}\hat n_{\bar i\bar\sigma}}-\braket{\hat n_{i\bar\sigma}\hat n_{\bar i\bar\sigma}\hat n_{\bar i\sigma}})\;,\\
    r_{i,4} &=& \braket{\hat n_{i\bar\sigma}\hat n_{\bar i\bar\sigma}\hat n_{\bar i\sigma}}\;,\\
    r_{i,5} &=& 2(\braket{\hat n_{\bar i\sigma}}-\braket{\hat n_{\bar i\sigma}\hat n_{\bar i\bar\sigma}} 
    -\braket{\hat n_{i\bar\sigma}\hat n_{\bar i\bar\sigma}} \nonumber\\
    && + \braket{\hat n_{i\bar\sigma}\hat n_{\bar i\bar\sigma}\hat n_{\bar i\sigma}})\;, \hspace{5ex}
    \\
    r_{i,6} &=& \braket{\hat n_{\bar i\sigma}\hat n_{\bar i\bar\sigma}}-\braket{\hat n_{i\bar\sigma}\hat n_{\bar i\bar\sigma}\hat n_{\bar i\sigma}}\;.
  \end{eqnarray}
  \label{eq_numerators}
\end{subequations}
\cref{eq_GF0} provides the structure of the one body GF as a sum of single poles, broadened by the imaginary part of the hybridization function $\Delta_i$.
In the wide-band limit (WBL) the hybridization function becomes constant and purely
imaginary, $\Delta_i=-i\gamma_{i}/2$, with $\gamma_i=\gamma_{i,L}+\gamma_{i,R}$. In the following we consider that $\gamma=\gamma_1=\gamma_2$. We emphasize that also the higher-order dot GFs have a
similar structure as \cref{eq_GF0} but in this case only a subset of the poles
of \cref{eq_pi} contribute, see \cref{app_Green_functions}.
The imaginary contribution in the denominators is obtained under the
approximation $[ \hat n_{i\sigma},\hat{\mathcal{H}}]=0$ and it formally justifies the
usual phenomenological procedure in which $\gamma$ is introduced as a decay
parameter in order to broaden the poles of the spectral function for different
impurity systems~\cite{ryndyk2009green,haug2008quantum,niu1999equation}.

The second step in our analytical EOM approach rests on the observation
that one can obtain the static $N$-body correlator from the GF
via~\cite{zubarev1960double,stefanucci2013nonequilibrium}
\begin{align}
	\label{eq_exp_values}
	&\braket{\hat{D}\,\hat{n}_{i\sigma}}=-\int f(\omega)\, \text{Im}\left(G_{D d_{i\sigma}}^{r}(\omega)\right)\;.
\end{align}
Since the GFs $G_{D d_{i\sigma}}^{r}(\omega)$ only depend
  {\it linearly} on the correlators, application of \cref{eq_exp_values} for
all GFs then leads to a linear system for the correlators. Importantly, 
the fourth-order correlator
$\braket{\hat n_{1\sigma}\hat n_{1\bar\sigma}\hat n_{2\sigma}\hat n_{2\bar\sigma}}$
does not explicitly appear in the resulting linear system, which thus in
principle consists of 15 equations for 15 independent correlators.

The size of the linear system can be reduced due to the spin symmetry in our
model (the on-site energies $v_i$ are independent of spin), which reduces the
number of independent correlators from 15 to 7 (see Appendix \ref{app_correlators}). Incidentally, in the method of rate equations
\cite{Beenakker:91,liul2023interferometry,mazal2019nonmonotonic}
one solves for the same number of independent variables, which in this case
are the probabilities for a certain distribution of occupation numbers in
the double dot. In fact, we expect the rate equation approach to be
physically equivalent to the EOM approach at our level of approximate
treatment of the coupling to the leads.

Another important simplification arises from the fact that all GFs
have the analytic structure of a sum of single poles in the complex
plane and therefore each integral arising on the r.h.s. of
\cref{eq_exp_values} has the same structure and can be evaluated
analytically~\cite{sobrino2021thermoelectric,bulka2004electronic}
\begin{align}
  \phi(p) & =  \int f(\omega)\frac{\gamma}{(\omega-p)^{2}+\frac{\gamma^{2}}{4}}\nonumber\\
  & =\frac{1}{2}-\frac{1}{\pi}\psi\left(\frac{1}{2}+\frac{\gamma/2+ip}{2\pi T} \right)\;,
            \label{eq_int_digamma}
\end{align}
where  $\psi(z)=\frac{d \log(\Gamma(z))}{dz}$ is the digamma function with
general complex argument $z$, and $\Gamma(z)$ is the gamma function. For
instance, the two-body correlator function
$\braket{\hat n_{i\bar\sigma}\hat n_{\bar i\bar\sigma}}$, can be expressed as

\begin{align}
  &\braket{\hat n_{i\bar\sigma}\hat n_{\bar i\bar\sigma}}=-\int  f(\omega)\text{Im}\left(\braket{\braket{\hat n_{\bar i\bar\sigma}\hat d_{i\bar\sigma}:\hat d^{\dagger}_{i\bar\sigma}}}\right)\nonumber\\
  &= \phi(p_{i,5}) \left(\braket{\hat n_{\bar i\sigma}} -
  \braket{\hat n_{\bar i\sigma}\hat n_{\bar i\bar\sigma}} -
  \braket{\hat n_{i\bar\sigma}\hat n_{\bar i\bar\sigma}} +
  \braket{\hat n_{i\bar\sigma}\hat n_{\bar i\bar\sigma}\hat n_{\bar i\sigma}} \right)
  \nonumber\\
  & \;\; + \phi(p_{i,6}) \left( \braket{\hat n_{\bar i\sigma}\hat n_{\bar i\bar\sigma}} -
  \braket{\hat n_{i\bar\sigma}\hat n_{\bar i\bar\sigma}\hat n_{\bar i\sigma}} \right)
  \nonumber\\
  & \;\; + \phi(p_{i,3}) \left( \braket{\hat n_{i\bar\sigma}\hat n_{\bar i\bar\sigma}} -
  \braket{\hat n_{i\bar\sigma}\hat n_{\bar i\bar\sigma}\hat n_{\bar i\sigma}} \right)
  \nonumber\\
  & \;\; + \phi(p_{i,4})
  \braket{\hat n_{i\bar\sigma}\hat n_{\bar i\bar\sigma}\hat n_{\bar i\sigma}} \;.
\end{align}
In other words, {\it all} coefficients in the linear system for the correlators
can be computed analytically. Of course, once these coefficients are known, the
resulting linear system can readily be solved numerically. However, in the
case of our double dot one can even solve the whole linear system analytically.
To this end, for the quantum dot with label $i$ we first solve the
subsystem for the higher-order correlators
$\braket{\hat n_{i\bar\sigma}\hat n_{\bar i\bar\sigma}}$,
$\braket{\hat n_{\bar i\sigma}\hat n_{\bar i\bar\sigma}}$,
$\braket{\hat n_{i\bar\sigma}\hat n_{\bar i\bar\sigma}\hat n_{\bar i\sigma}}$,
and $\braket{\hat n_{i\sigma}\hat n_{i\bar\sigma}\hat n_{\bar i\sigma}}$
and express them in terms of the occupations $\braket{\hat n_{i\sigma}}=n_i/2$,
and $\braket{\hat n_{\bar i\sigma}}=n_{\bar i}/2 $. Subsequently, we solve the
resulting system for the two local occupations (see \cref{app_correlators} for the full derivation).

The higher-order correlators are expressed in terms of the occupations as
\begin{subequations}
	\begin{align}
		&\braket{\hat n_{\bar i\sigma}\hat n_{\bar i\bar\sigma}}=\tau_{i,1}\braket{\hat n_{\bar i\sigma}}\;,\\
		&\braket{\hat n_{i\bar\sigma}\hat n_{\bar i\bar\sigma}}=\tau_{i,2}\braket{\hat n_{\bar i\sigma}}\;,\\
		&	\braket{\hat n_{i\bar\sigma}\hat n_{\bar i\bar\sigma}\hat n_{\bar i\sigma}}=\tau_{i,3}\braket{\hat n_{\bar i\sigma}}\;,\\
		&	\braket{\hat n_{i\sigma}\hat n_{i\bar\sigma}\hat n_{\bar i\sigma}} =\tau_{i,4}\braket{\hat n_{\bar i\sigma}}\;,
	\end{align}
	\label{eq_solution_two_body}
\end{subequations}
$\!\!$where the parameters $\tau_{i,j}$ are defined in \cref{eq_iota_coeff}
and are fully given in terms of the model parameters.
Substituting \cref{eq_solution_two_body} into \cref{eq_numerators} allows us to
express the residues of the single-particle GF in terms of the orbital
occupations
\begin{subequations}
	\begin{align}
	r_{i,1} =& 1-n_i/2+\left(\tau_{i,1}+2\tau_{i,2}-\tau_{i,3}-2\right)n_{\bar i}/2\;,\\
	r_{i,2} =& n_i/2+\left(\tau_{i,3}-2\tau_{i,2}\right)n_{\bar i}/2\;,\\
	r_{i,3} =&\left(\tau_{i,2}-\tau_{i,3}\right)n_{\bar i}\;,\\
	r_{i,4} =&\tau_{i,3} n_{\bar i}/2\;,\\
	r_{i,5} =&\left(1-\tau_{i,1}-\tau_{i,2}+\tau_{i,3}\right)n_{\bar i} \;,\\
	r_{i,6} =&\left(\tau_{i,1}-\tau_{i,3}\right)n_{\bar i}/2\;.
	\end{align}
	\label{eq_numerators_one_body}
\end{subequations}
Combining \cref{eq_exp_values1,eq_GF0,eq_numerators_one_body} one then
  arrives at a $2 \times2$ linear system for the local occupations which
  is readily solved analytically, see \cref{app_correlators}.
The resulting local occupations are then given by:
\begin{align}
n_i=2\frac{\phi(v_i)\eta_{\bar i\bar i}-\phi(v_{\bar i})\eta_{i\bar i}}{\eta_{11}\eta_{22}-\eta_{12}\eta_{21}}\;,
\label{eq_local_occupations}
\end{align}
where 
\begin{subequations}
	\begin{align}
		\eta_{ii}=&1+\phi(p_{i,1})-\phi(p_{i,2})\;,\\
		\eta_{i\bar i}=& \phi(p_{i,1})\left(2+\tau_{i,3}-\tau_{i,1}-2\tau_{i,2} \right) + \phi(p_{i,2})\left(2\tau_{i,2}-\tau_{i,4} \right)\nonumber\\
		+& 2\phi(p_{i,3}) \left(\tau_{i,3}-\tau_{i,2} \right)-\tau_{i,3} \phi(p_{i,4})\nonumber\\
		+& 2\phi(p_{i,5})\left(\tau_{i,1}+\tau_{i,2}-\tau_{i,3}-1 \right) + \phi(p_{i,6})\left(\tau_{i,3}-\tau_{i,1} \right)\;.
	\end{align}
	\label{eq_eta_params}
\end{subequations}
\cref{eq_GF0,eq_numerators_one_body,eq_local_occupations} represent the main results of this work, providing analytical expressions for the evaluation of the
Green's functions and the orbital occupations as functions of the model
parameters. 

In the uncontacted limit $\gamma\to0$, our method
involves no approximations and treats interactions exactly. The imaginary part
of the resulting GF (\cref{eq_GF0}) is proportional to a sum of
delta functions at frequencies coinciding with the poles $p_{i,j}$ which allows us
to express the correlators in terms of Fermi functions, $\phi(p)=f(p)$.
The results in this limit are in complete agreement with the grand canonical
ensemble (GCE) as it should be.

A common alternative approach to treat the coupling dependence consists in
truncating the  mixed GFs hierarchy at the level of the two-particle GF,
considering the Hubbard-I \cite{hubbard1963electron} decoupling $\braket{\braket{\hat{n}_{i'\sigma'}\hat{c}_{ik\sigma }:d^{\dagger}_{i\sigma}}}\approx\braket{\hat{n}_{i'\sigma'}}\braket{\braket{\hat{c}_{ik\sigma }:d^{\dagger}_{i\sigma}}}$ and treating the inter-dot Coulomb interaction within the Hartree approximation. This method (denoted as EOM-H in the following) has already been explored in the literature \cite{sierra2016interactions} and the resulting one-particle GF is 
\begin{align}
G_{i\sigma}^r(\omega)&=\frac{1-n_i/2}{\omega-p_{i,1}-U_{12}n_{\bar i}-\Delta_i \left(1+\frac{U_i \,n_i/2}{\omega-p_{i,2}}\right)}\nonumber\\
&+\frac{n_i/2}{\omega-p_{i,2}-U_{12}n_{\bar i}-\Delta_i \left(1-\frac{U_i(1-n_i/2)}{\omega-p_{i,1}}\right)}\;.
\label{eq_G_Hartree}
\end{align}
Note that since the spectral function related to \cref{eq_G_Hartree} can not
be written as a sum of Lorentzian functions, the resulting integrals
  for obtaining the occupations have to be solved numerically. Furthermore, the
  resulting system of equations is nonlinear in the occupations which
cannot be solved analytically and requires a numerical approach.

\section{Results}
\label{results}

In this section,  we apply our analytical EOM approach to compute spectral functions and orbital occupations
for the DQD in different regimes defined by the interaction parameters (c.f. Ref.~\onlinecite{sobrino2020exchange}).
We benchmark the results with two well-known numerical methods, namely the non-crossing approximation (NCA)~\cite{coleman1984nca,bickers1987}
and the hierarchical equations of motion (HEOM) approach~\cite{tanimura1990nonperturbative,yan2004hierarchical,xu2005exact}.
A Python code implementing our EOM approach for the DQD is publicly available at GitHub~\cite{github_DQD}.

\begin{figure*}
  \includegraphics[width=\linewidth]{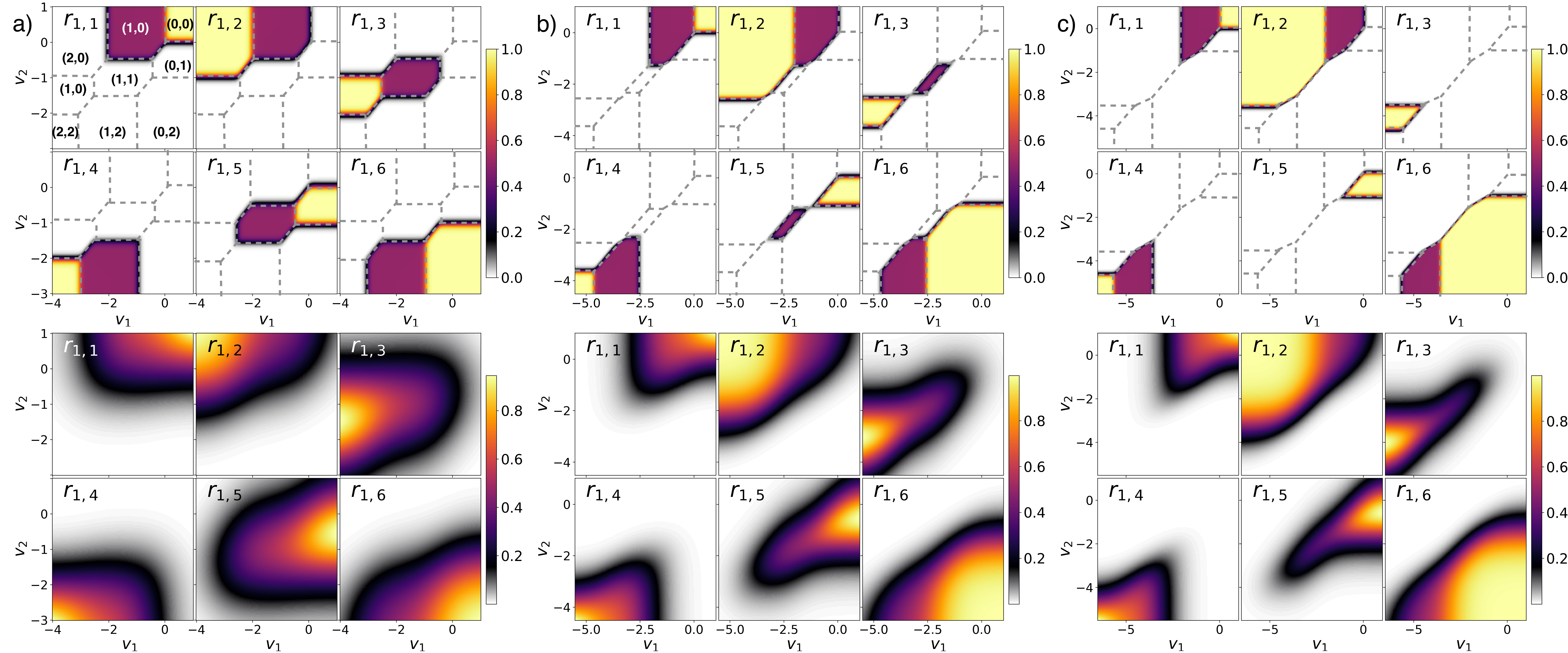}
  \caption{\label{fig_1}
    Green's function residues $r_{1,j} $ of \cref{eq_numerators_one_body} as a function of the gate potentials for different
    combinations of the interaction parameters at \(T=0.05\) (top panels) and  $T=0.5$ (bottom panels) in the limit \(\gamma \to 0\). The Coulomb interactions are: (left)
    \(U_1 = 2\) and \(U_{12} = 0.5\), (center) \(U_1 = 2\) and \(U_{12} = 1.3\), and (right) \(U_1 = 2\) and \(U_{12} = 1.8\). Energies given in units of $U_{2}$.
  }
\end{figure*}

In the following we consider the three different regimes of interaction parameters $U_1$, $U_2$, and $U_{12}$ identified in
Ref.~\onlinecite{sobrino2020exchange}: In Regime I the inter-dot interaction is smaller than either of the intra-dot interactions.
Without loss of generality we may also assume $U_2\le{U_1}$, and hence $U_{12}\le{U_2}\le{U_1}$. Regime I is the most common (or natural)
case, since in physical systems usually the inter-dot (or inter-orbital) interaction will be smaller than the intra-dot (or intra-orbital)
ones. On the other hand, Regimes II ($(U_1+U_2)/2>U_{12}>U_2$) and III ($U_{12}> (U_1+U_2)/2 >U_2$) where $U_{12}$ is larger than at least one
of the on-site interactions $U_i$ could be relevant in specific setups or for effective model Hamiltonians.

In order to better understand the structure of the GF given by \cref{eq_GF0},
in \cref{fig_1} we plot the residues given by
\cref{eq_numerators_one_body}  for QD $i=1$ as functions of the gate
potentials $v_i$ of both QDs for each of the three aforementioned regimes,
together with so-called stability diagrams structure (shown as dashed gray lines) in the
limit of low coupling and both low (top panels) and higher (bottom panels) temperatures (on the scale of the interactions).
A stability diagram is a map of the dot occupation numbers $\{n_1,n_2\}$ as a
function of the gate potentials $\{v_1,v_2\}$. It thus shows the regions where
the occupation numbers are stable, i.e., do not fluctuate in the limit of low
temperature and weak coupling to the reservoirs.
Specifically in this limit and for the case of only density-density
interactions as in our case, the occupations in the different regions
are stable at integer values.
As explained in detail in Ref.~\onlinecite{sobrino2020exchange}, each regime
is characterized by a distinctive stability diagram.

Regime I, shown in \cref{fig_1}(a), is characterized by relatively independent
behavior of both dots and a wide central region of occupations $(1,1)$.
In Regime II, shown in \cref{fig_1}(b), the width of the $(1,1)$ region in the
stability diagram is notably reduced, leading to increased regions of the $(1,2)$
and $(2,1)$ occupation. Finally, in Regime III, shown in \cref{fig_1}(c), the $(1,1)$ region of occupations
is not present anymore, and the regions $(0,1)$ and $(2,0)$ as well as the regions
$(1,0)$ and $(0,2)$ become adjacent. 

We now focus on a given regime, i.e., fixed interactions
$U_1$, $U_2$ and $U_{12}$, and a given region in the $v_1-v_2$ plane. By
inspection of the top panels of \cref{fig_1} (low
  coupling and low temperature), we realize that for each given region there
are two situations possible: (i) there exists only one non-vanishing residue
with value 1 or (ii) there exist two non-vanishing residues both with value
0.5, i.e., the local spectral function (on dot 1) has either one or two
poles. In the first case the single pole is at an electron removal or
electron addition energy for an electron at the given site. In the case of
two poles, the corresponding site is half filled and one can both add and
remove an electron to/from that site. Note that both poles have equal
spectral weight, i.e., equal residues, because the dot Hamiltonian only
contains density-density interactions. As the temperature is increased,
eigenstates different from the ground state start to contribute to the
ensemble of the dot, and thus more than two poles may contribute to the
local spectral function, as can be seen in  the bottom panels of \cref{fig_1}
(higher temperature).

\begin{figure}[ht]
  \includegraphics[width=\linewidth]{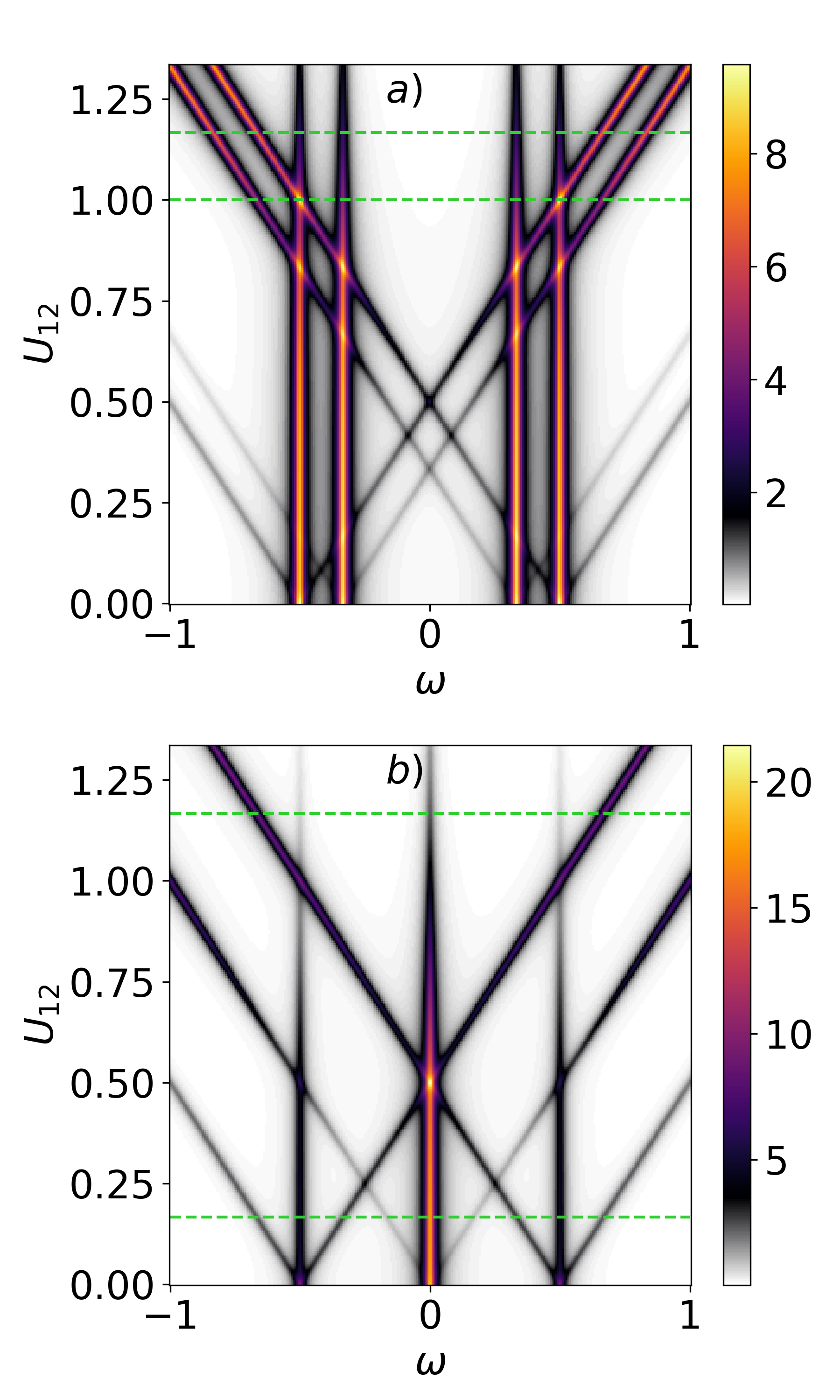}
  \caption{Spectral function of the double quantum dot from
    \cref{eq_GF0,eq_numerators_one_body} for \(T = 1/6\), \(\gamma = 1/30\) at the particle-hole
    symmetric point \(v_i = -U_{i}/2 - U_{12}\).
    (a) \(U_1 = 2/3\) and (b) \(U_1 = 0\). Energies in units of $U_{2}$.}
  \label{fig_2}
\end{figure}

In \cref{fig_2}, the total spectral function \(A(\omega) = -\text{Im}[G_{1}^{r}(\omega) + G_{2}^{r}(\omega)] / \pi\)
obtained from the analytical EOM, is plotted as a function of the inter-dot Coulomb repulsion \(U_{12}\) at the
particle-hole symmetric point \(v_i = -U_{i}/2 - U_{12}\), with \(T/\gamma = 5\), \(U_2 = 1\), and (a) \(U_1 = 2/3\)
and (b) \(U_1 = 0\). The main structure of the spectral function consists of a series of intersecting vertical and
diagonal lines. The vertical lines are located at \(\omega = \pm U_i/2\) while
the diagonals follow from \(\pm\omega = \pm U_i/2 - U_{12}\)
consistent with the position of the poles \cref{eq_pi} for
\(v_i = -U_{i}/2 - U_{12}\). 

An important observation is that for small \(U_{12}\) values, almost all the
spectral weight is carried by the vertical lines. As \(U_{12}\) increases, the
spectral weight shifts to the diagonal lines. 
This occurs at the transitions to regime II (at (panel a) \(U_{12} = 2/3\)
and to regime III (at  (panel a) \(U_{12} = 5/6\) and (panel b)
\(U_{12} = 1/2\)). The lines with less spectral weights are
due to excited states and their weight decreases in the limit of both
$T\to 0$ and $\gamma \to 0$ (not shown).

\begin{figure}[ht]
  \includegraphics[width=\linewidth]{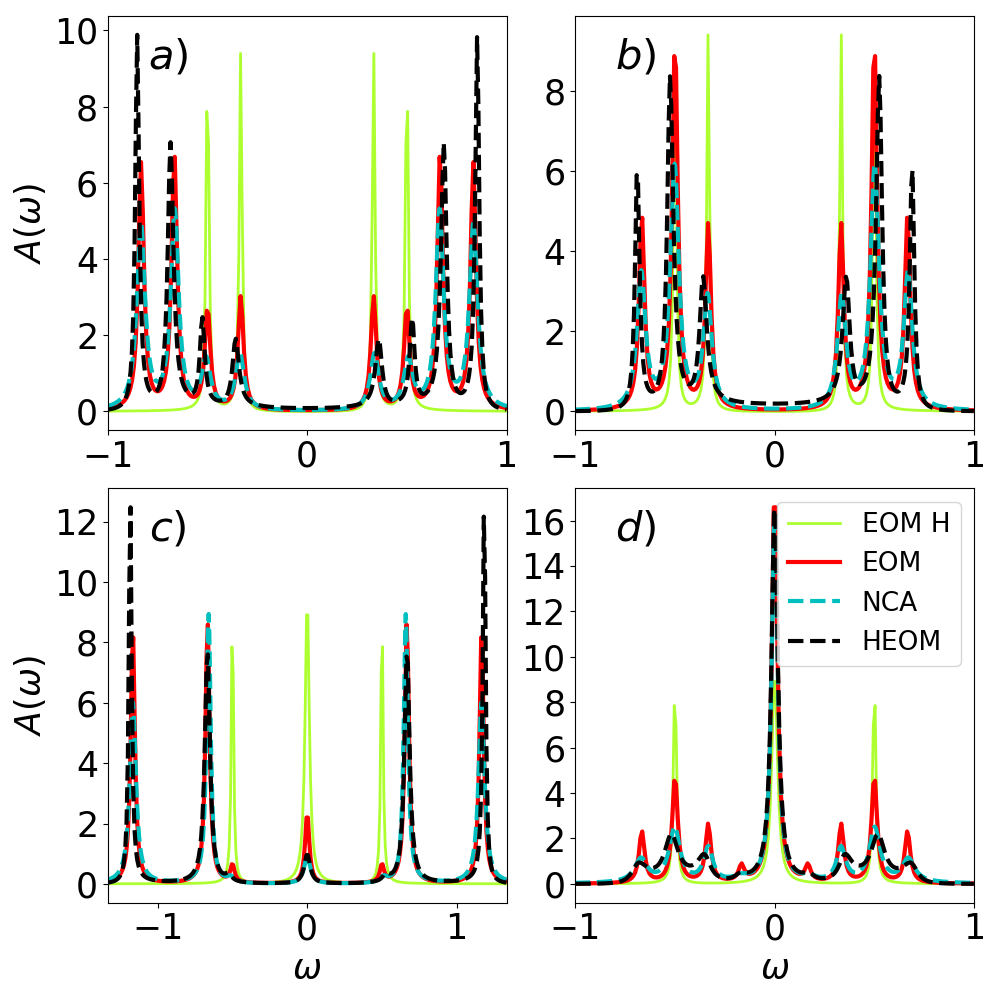}
  \caption{\label{fig_3}
    Spectral function of the double quantum dot at \(T = 1/6\), \(\gamma = 1/30\) and at the particle-hole symmetric point
    \(v_i = -U_{i}/2 - U_{12}\). In (a) \(U_1 = 2/3\), \(U_{12} = 7/6\), (b) \(U_1 = 2/3\), \(U_{12} = 1\), (c) \(U_1 = 0\), \(U_{12} = 7/6\), and (d)
    \(U_1 = 0\), \(U_{12} = 1/6\). The different lines correspond to the EOM Hartree approximation (EOM-H) (divided by 2 for representation clarity),
    the analytical EOM of \cref{eq_GF0,eq_numerators_one_body}, and the numerical results from NCA and HEOM.
Energies in units of $U_{2}$.  }
\end{figure}

In \cref{fig_3}, we compare our results for the total spectral function (EOM)
for those values of $U_{12}$ marked by the green dashed lines in
\cref{fig_2} against three other methods: the Hubbard I + Hartree decoupling
scheme of \cref{eq_G_Hartree} (EOM-H),  NCA, and HEOM for tier level \(L = 3\).
At small \(U_{12}\) values, the EOM-H approach captures the general distribution of the
spectral functions (corresponding to the vertical lines in \cref{fig_2}) in comparison to NCA
and HEOM. However, as \(U_{12}\) increases, EOM-H
fails to capture the correct spectral weights of the different peaks,
finding that the vertical lines still carry all the spectral weight (instead
of the diagonal ones). As shown, the analytical EOM accurately reproduces the
position of all the peaks and correctly captures the spectral weight distributions for all values of \(U_{12}\).

\begin{figure}[ht]
  \includegraphics[width=\linewidth]{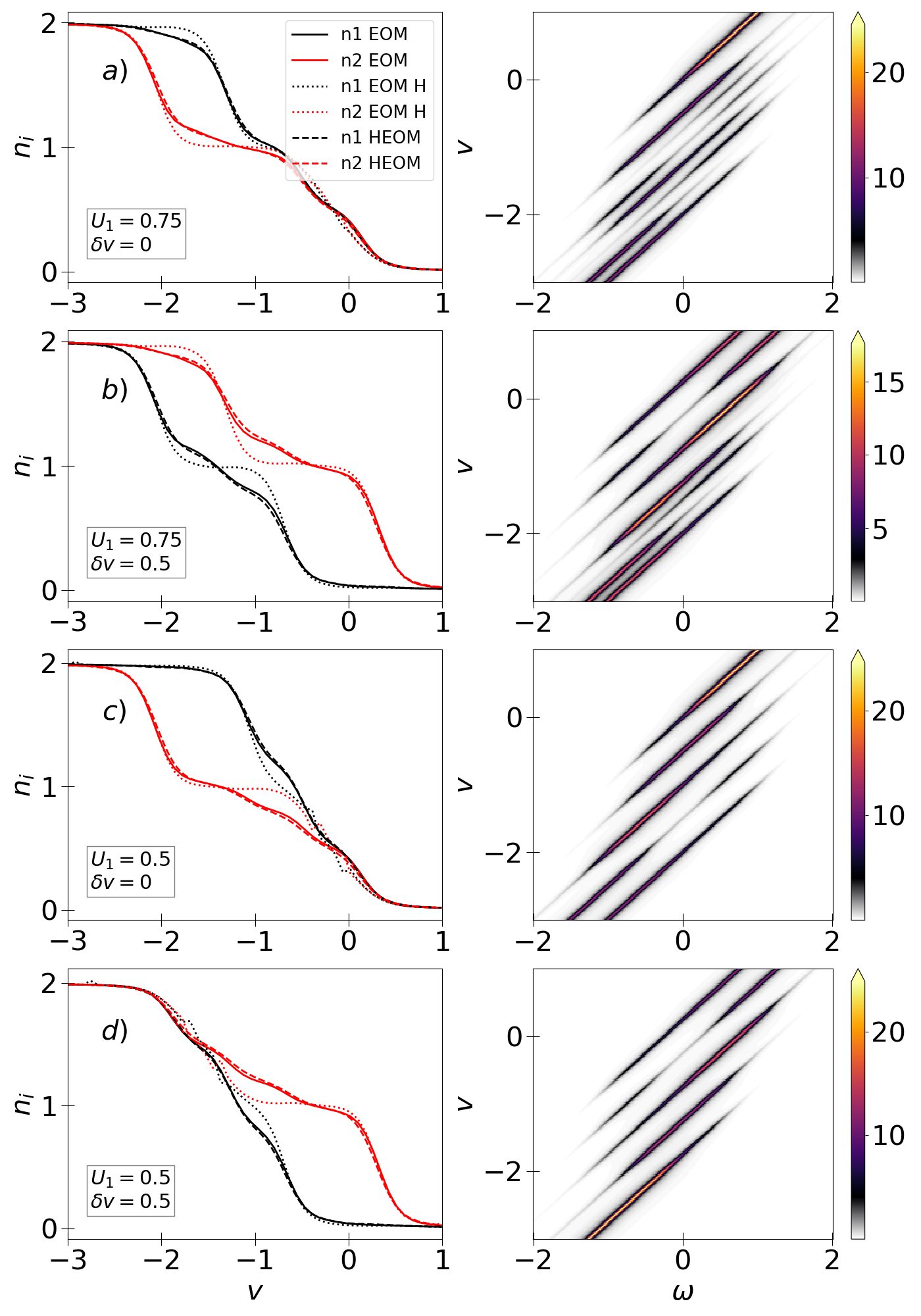}
  \caption{\label{fig_4}Local occupations (left) and spectral functions (right) as function of the average gate level $v = (v_1 + v_2)/2$ for $U_{12}=0.5$, $T=0.1$, and $\gamma=0.05$.
    The other interaction  and parameters are (a) $U_1 = 0.75$ and $\delta v = 0$, (b) $U_1 = 0.75$ and $\delta v = 0.5$, (c) $U_1 = 0.5$ and $\delta v = 0$, and (d) $U_1 = 0.5$ and $\delta v = 0.5$. Energies in units of $U_{2}$.}
\end{figure}

Finally, in the left columns of \cref{fig_4} we present a comparison of the
local occupations as a function of the average gate level $v = (v_1 + v_2)/2$
between our analytical EOM approach, the EOM-H approximation and the HEOM
method for different interactions and $\delta v = v_1 - v_2$, at a fixed
$T= 2\gamma=0.1$.
In panel a)  with $U_1 = 0.75$, the local occupations are degenerate  up to half occupation at $v \sim -0.5$. Then, $n_1$ evolves faster to full occupation due to the smaller intra-dot Coulomb repulsion. In panel b) where a finite energy difference is applied between the gate levels, $\delta v=v_1-v_2= 0.5$, the occupation of site one is shifted to lower gates, with two plateaus close to half filling being the main features of the densities.  In panel c)  where $U_1 = 0.5$
and $\delta v = 0$, both $n_i$ evolve together as the gate is decreased up to quarter occupation at $v \sim0$. As in a), then $n_1$ evolves faster to full occupation. Finally, in in panel d), the local occupation of site one is shifted to lower gate levels $v$. In all the situations studied with $T/\gamma > 1$, the agreement between the EOM approach and the HEOM is excellent, with our analytical EOM approach accurately capturing the local step structures due to the interactions and the corresponding slopes governed by the combination of temperature and coupling strength.  On the other hand, although the EOM-H approximation
captures the general trend of the local occupations, it fails to accurately predict some local transitions. In particular, the EOM-H approximation does not correctly capture the new plateaus at non-integer values of the  local occupations observed in the EOM and HEOM results. This discrepancy highlights the limitation of the EOM-H method in dealing with the  inter-dot Coulomb repulsion, not recovering correctly the GCE limit (not shown).
	
The right panels in \cref{fig_4} depict the spectral functions corresponding
to each case on the left as a function of the average gate $v$.
As expected, all non-vanishing spectral features depend linearly on $v$ due to the
positions of the poles given by $\omega = p_{i,j}$. 
The weight of each contribution (given by the residues) varies as the
interaction changes, but at full occupation, all the spectral weight is
carried by $r_{i,4}$ since in this case one can only remove an electron
with a removal energy given by $p_{i,4}$. In the empty dot case, all the weight
is carried by the residue $r_{i,1}$, since now one can only add an
electron with an addition energy given by $p_{i,1}$.

\section{Conclusions}
\label{conclus}

In this paper, we have introduced a fully analytical EOM approach to analyze the spectral
properties and orbital occupations of an interacting DQD system connected to leads.
A critical aspect of our approach is the systematic closure of the hierarchy of higher-order
GFs, which arises from a consistent approximate treatment of the coupling
  to the leads for the GFs at all orders.
This closure is essential to map the EOMs for the GFs onto a linear system
  for the density correlators, a mapping that can be applied to any number of levels. For
  our double dot system it allows us to derive tractable 
analytical expressions for the one-body GF in terms of the local occupations and all the model parameters.
In the low coupling limit, our method correctly recovers the
results of the grand canonical ensemble, further validating the approach.

We compared our analytical results with
those obtained from the NCA and the HEOM methods. Our approach accurately reproduces the spectral peak positions and weight distributions, demonstrating its effectiveness in capturing the essential physics of the strongly correlated electron system in the Coulomb blockade regime. The agreement with these established numerical techniques underscores the reliability of our approach.

The analytical EOM approach offers substantial advantages, including explicit functional dependencies that provide  insights into the physical mechanisms governing the system. Furthermore, the computational efficiency of our analytical method enables a cheap
exploration of different parameter regimes of the system and the study of properties such as the transport coefficients. Additionally, this computational efficiency may be particularly beneficial for, e.g., density functional theory and its extensions, where
knowledge of the many-body spectral function in analytic form can be of crucial importance for the construction of the exchange-correlation potentials by reverse engineering.\cite{StefanucciKurth:15,kurth2017transport,JacobKurth:18,sobrino2019steady,sobrino2023thermoelectric}.

\appendix

\section{Approximate treatment of coupling to leads}
\label{app_derivation_approx}

The purpose of the present Appendix is to treat the coupling to the leads
directly from the EOM and highlight the approximations which lead to the contribution
of the standard embedding self energy of \cref{eq_Delta} in both the single-particle and
higher order GFs (which in the WBL leads to simple broadening of all poles of the GFs by the
wide-band parameter $\gamma$ used in Sec.~\ref{model_eom}).

In Sec.~\ref{model_eom} we have defined the Green function
$\langle \langle \hat{D} \hat{d}_{i \sigma} : \hat{d}_{i \sigma}^{\dagger}
\rangle \rangle_{\omega}$ with the operator
$\hat{D} = \hat{n}_{i_1 \sigma_1} \hat{n}_{i_2 \sigma_2} \ldots \hat{n}_{i_N \sigma_N}$
being a product of $N$ density operators with $(i_k \sigma_k) \neq (i \sigma)$
for all $k \in \{1, \ldots,N\}$. In addition, for $N=0$ we define
$\hat{D} = \hat{\mathbbm{1}}$ as the unit operator. We note that
$\hat{d}_{i \sigma}$ commutes with $\hat{D}$ for all $N$, i.e.,
$[ \hat{D}, \hat{d}_{i \sigma} ] = [ \hat{D}, \hat{d}_{i \sigma}^{\dagger}] = 0$.

After straightforward, if somewhat tedious calculations one obtains
the EOM for $\langle \langle \hat{D} \hat{d}_{i \sigma} :
\hat{d}_{i \sigma}^{\dagger} \rangle \rangle_{\omega}$ which reads
\begin{eqnarray}
\lefteqn{
  (\omega - v_i) \langle \langle \hat{D} \hat{d}_{i \sigma} :
  \hat{d}_{i \sigma}^{\dagger} \rangle \rangle_{\omega}} \nonumber\\
&=& \langle \hat{D} \rangle
+ U_i \langle \langle \hat{D} \hat{n}_{i \bar{\sigma}} \hat{d}_{i \sigma} :
\hat{d}_{i \sigma}^{\dagger} \rangle \rangle_{\omega} \nonumber\\
&& + \sum_{j \sigma'} U_{ij} \langle \langle \hat{D} \hat{n}_{j \sigma'}
\hat{d}_{i \sigma} : \hat{d}_{i \sigma}^{\dagger} \rangle \rangle_{\omega} \nonumber\\
&& + \sum_{k} V_{\alpha k i}^* \langle \langle \hat{D}_N \hat{c}_{\alpha k i \sigma} :
\hat{d}_{i \sigma}^{\dagger} \rangle \rangle_{\omega} \nonumber\\
&& + \sum_{j \alpha k \sigma'} V_{\alpha k j}^* \langle \langle [ \hat{d}_{j \sigma'}^{\dagger},
  \hat{D} ] \hat{d}_{i \sigma} \hat{c}_{\alpha k j \sigma'} :
\hat{d}_{i \sigma}^{\dagger} \rangle \rangle_{\omega} \nonumber\\
&& + \sum_{j \alpha k \sigma'} V_{\alpha k j} \langle \langle \hat{c}_{\alpha k j\sigma'}^{\dagger}
    [ \hat{d}_{j \sigma'}, \hat{D} ] \hat{d}_{i \sigma} :
    \hat{d}_{i \sigma}^{\dagger} \rangle \rangle_{\omega}\;.
    \label{eom_Dd}
\end{eqnarray}
Similarly, the EOM for the mixed dot-lead GF $\langle \langle \hat{D} \hat{c}_{\alpha k i\sigma} :
\hat{d}_{i \sigma}^{\dagger} \rangle \rangle_{\omega}$ reads
\begin{eqnarray}
\lefteqn{(\omega - \epsilon_{\alpha k i})
  \langle \langle \hat{D} \hat{c}_{\alpha k i\sigma} :
  \hat{d}_{i \sigma}^{\dagger} \rangle \rangle_{\omega}}\nonumber\\
&=& V_{\alpha k i} \langle \langle \hat{D} \hat{d}_{i \sigma} :
\hat{d}_{i \sigma}^{\dagger} \rangle \rangle_{\omega} \nonumber\\
&& + \sum_{j \alpha' q \sigma'} V_{\alpha' q j}^* \langle \langle [ \hat{D}  ,
  \hat{d}_{j \sigma'}^{\dagger} ] \hat{c}_{\alpha' q j \sigma'} \hat{c}_{\alpha k i \sigma} :
\hat{d}_{i \sigma}^{\dagger} \rangle \rangle_{\omega} \nonumber\\
&& + \sum_{j \alpha' q \sigma'} V_{\alpha'  q j} \langle \langle \hat{c}_{\alpha' q j \sigma'}^{\dagger}
    \hat{c}_{\alpha k i\sigma} [ \hat{d}_{j \sigma'}, \hat{D} ] :
    \hat{d}_{i \sigma}^{\dagger} \rangle \rangle_{\omega}\;.
    \label{eom_Dc}
\end{eqnarray}
If in Eqs.~(\ref{eom_Dd}) and (\ref{eom_Dc}) we make the approximations
\begin{eqnarray}
\langle \langle [ \hat{d}_{j \sigma'}^{\dagger},
  \hat{D} ] \hat{d}_{i \sigma} \hat{c}_{\alpha k i \sigma'} :
\hat{d}_{i \sigma}^{\dagger} \rangle \rangle_{\omega}  \approx 0\;, \nonumber\\
\langle \langle \hat{c}_{\alpha k i \sigma'}^{\dagger}
    [ \hat{d}_{j \sigma'}, \hat{D} ] \hat{d}_{i \sigma} :
    \hat{d}_{i \sigma}^{\dagger} \rangle \rangle_{\omega} \approx 0 \;,\nonumber\\
    \langle \langle [ \hat{D} ,\hat{d}_{j \sigma'}^{\dagger} ]
    \hat{c}_{\alpha' q j \sigma'} \hat{c}_{\alpha k i \sigma}
    :\hat{d}_{i \sigma}^{\dagger} \rangle \rangle_{\omega}
    \approx 0\;, \nonumber \\
    \langle \langle \hat{c}_{\alpha' q j \sigma'}^{\dagger}
    \hat{c}_{\alpha k i \sigma} [ \hat{d}_{j \sigma'}, \hat{D} ] :
    \hat{d}_{i \sigma}^{\dagger} \rangle \rangle_{\omega} \approx 0\;,
    \label{gf_approx_zero}
\end{eqnarray}
 insert the resulting Eq.~(\ref{eom_Dc}) into (\ref{eom_Dd})
and obtain
\begin{eqnarray}
\lefteqn{
  (\omega - v_i - \Delta_{i}(\omega))
  \langle \langle \hat{D} \hat{d}_{i \sigma} :
  \hat{d}_{i \sigma}^{\dagger} \rangle \rangle_{\omega}} \nonumber\\
&=& \langle \hat{D} \rangle
+ U_i \langle \langle \hat{D} \hat{n}_{i \bar{\sigma}} \hat{d}_{i \sigma} :
\hat{d}_{i \sigma}^{\dagger} \rangle \rangle_{\omega} \nonumber\\
&& + \sum_{j \sigma'} U_{ij} \langle \langle \hat{D} \hat{n}_{j \sigma'}
\hat{d}_{i \sigma} : \hat{d}_{i \sigma}^{\dagger} \rangle \rangle_{\omega}\;.
\label{eom_approx}
\end{eqnarray}
We note that for the single-particle GF,
the approximation of Eq.~(\ref{gf_approx_zero}) becomes exact and we
obtain \cref{eq_EOM_3}. 

\section{Green functions}
\label{app_Green_functions}
Since \cref{eq_H} is time-independent, the equation of motion of a general Green's function $\braket{\braket{\hat B(t):\hat d_{i\sigma}^{\dagger}(0)}}$ can be expressed in the frequency domain through the relation \cite{hewson1966theory}
\begin{equation}
\omega\braket{\braket{\hat B:\hat d_{i\sigma}^{\dagger}}}_{\omega} = \braket{\{\hat B,\hat d_{i\sigma}^{\dagger}\}}+\braket{\braket{[\hat B,\hat{\mathcal{H}}]:\hat d_{i\sigma}^{\dagger}}}_{\omega}
\label{eq_motion_corr_2}\;.
\end{equation}
The one body GF $\braket{\braket{\hat d_{i\sigma}:\hat d_{i\sigma}^{\dagger}}}_{\omega}$ is then derived from 
\begin{equation}
  \omega\braket{\braket{\hat d_{i\sigma}:\hat d_{i\sigma}^{\dagger}}}_{\omega} = \braket{\{\hat d_{i\sigma},\hat d_{i\sigma}^{\dagger}\}}+\braket{\braket{[\hat d_{i\sigma},\hat{\mathcal{H}}]:\hat d_{i\sigma}^{\dagger}}}_{\omega}\;,
  \label{eq_motion_G}
\end{equation}
with the anti commutation relations for the creation and annihilation operators	$\{\hat d_{i\sigma},\hat d_{j\sigma}^{\dagger}\}=\delta_{ij}$, $\{\hat d_{i\sigma},\hat d_{j\sigma}\}=\{\hat d_{i\sigma}^{\dagger},d_{j\sigma}^{\dagger}\}=0$. The commutator of the annihilation operators in the dots  of the DQD Hamiltonian follow
\begin{gather}
	[\hat d_{i\sigma},\hat{\mathcal{H}}]=\left(v_i+U_{i}\hat n_{i\bar \sigma}+U_{12}\hat n_{\bar i}\right)\hat d_{i\sigma}+\sum_{\alpha k}V^{*}_{\alpha k i}\hat c_{\alpha k i  \sigma}\;.
	\label{eq_commutator}
\end{gather}
Using Eq.~(\ref{eom_approx}) for $\hat{D}= \hat{\mathbbm{1}}$, the equation of motion for the
single-particle GF reads
\begin{align}
  \braket{\braket{\hat d_{i\sigma}:\hat d^{\dagger}_{i\sigma}}}_{\omega}
  &=\frac{1}{\omega-v_i-\Delta_{i}}+\frac{U_{i} \braket{\braket{\hat n_{i\bar\sigma}\hat d_{i\sigma}:\hat d^{\dagger}_{i\sigma}}}_{\omega}}{\omega-v_i-\Delta_{i}}\nonumber\\&+\frac{U_{12}}{\omega-v_i-\Delta_{i}} \sum_{\sigma'}\braket{\braket{\hat n_{\bar i \sigma'}\hat d_{i\sigma}:\hat d^{\dagger}_{i\sigma}}}_{\omega}\;.
\label{eq_GF_one_body}
\end{align}
The two-body GF $\braket{\braket{\hat n_{i\bar\sigma}\hat d_{i\sigma}:\hat d^{\dagger}_{i\sigma}}}_{\omega}$ follows from the evaluation of Eq.~(\ref{eom_approx}) with $\hat D =\hat n_{i\bar\sigma} $
which gives
\begin{gather}
	\braket{\braket{\hat n_{i\bar\sigma}\hat d_{i\sigma}:\hat d^{\dagger}_{i\sigma}}}_{\omega}
	= \frac{\braket{\hat n_{ i\bar\sigma}}}{\omega-v_i-U_i-\Delta_i}+\nonumber\\
        \sum_{\sigma'} \frac{U_{12}}{\omega-v_i-U_i-\Delta_i} 
        \braket{\braket{\hat n_{i\bar\sigma}\hat n_{\bar i \sigma'}
            \hat d_{i\sigma}:\hat d^{\dagger}_{i\sigma}}}_{\omega}\;,
	\label{eq_corr_two_body_1}
\end{gather}
while the other two two-particle GFs are
\begin{gather}
	\braket{\braket{\hat n_{\bar i \sigma}\hat d_{i\sigma}:\hat d^{\dagger}_{i\sigma}}}_{\omega}
	= \frac{\braket{\hat n_{\bar i \sigma}}}{\omega-v_i-U_{12}-\Delta_i}+\nonumber\\
        \frac{U_i \braket{\braket{\hat n_{i\bar\sigma}\hat n_{\bar i \sigma}\hat d_{i\sigma}:\hat d^{\dagger}_{i\sigma}}}_{\omega}}{\omega-v_i-U_{12}-\Delta_i} 
        +
        \frac{U_{12} \braket{\braket{\hat n_{\bar i \sigma}\hat n_{\bar i \bar\sigma}\hat d_{i\sigma}:\hat d^{\dagger}_{i\sigma}}}_{\omega}}{\omega-v_i-U_{12}-\Delta_i} \;,
	\label{eq_corr_two_body_2}
\end{gather}
and
\begin{gather}
	\braket{\braket{\hat n_{\bar i \bar \sigma}\hat d_{i\sigma}:\hat d^{\dagger}_{i\sigma}}}_{\omega}
	= \frac{\braket{\hat n_{\bar i \bar \sigma}}}{(\omega-v_i-U_{12}-\Delta_i)}+\nonumber\\
        \frac{U_i \braket{\braket{\hat n_{i\bar\sigma}\hat n_{\bar i \sigma}\hat d_{i\sigma}:\hat d^{\dagger}_{i\sigma}}}_{\omega}}{\omega-v_i-U_{12}-\Delta_i} 
        +
        \frac{U_{12} \braket{\braket{\hat n_{\bar i \sigma}\hat n_{\bar i \bar\sigma}\hat d_{i\sigma}:\hat d^{\dagger}_{i\sigma}}}_{\omega}}{\omega-v_i-U_{12}-\Delta_i} \;.
	\label{eq_corr_two_body_3}
\end{gather}
Similarly, for the three three-body GFs we obtain
\begin{gather}
\braket{\braket{\hat n_{i\bar\sigma}\hat n_{\bar i \sigma }\hat d_{i\sigma}:\hat d^{\dagger}_{i\sigma}}}_{\omega}=\frac{\braket{\hat n_{i\bar\sigma}\hat n_{\bar i\sigma}}}{\omega-v_i - U_i-U_{12}-\Delta_i}\nonumber\\
+\frac{U_{12}\braket{\braket{\hat n_{i\bar\sigma}n_{\bar i\sigma}\hat n_{\bar i\bar \sigma}\hat d_{i\sigma}:\hat d^{\dagger}_{i\sigma}}}_{\omega}}{\omega-v_i-U_i-U_{12}-\Delta_i} \;,
\label{eq_corr_three_body_1}
\end{gather}
\begin{gather}
\braket{\braket{\hat n_{\bar i\bar\sigma}\hat n_{\bar i \sigma }\hat d_{i\sigma}:\hat d^{\dagger}_{i\sigma}}}_{\omega}=\frac{\braket{\hat n_{\bar i\bar\sigma}\hat n_{\bar i\sigma}}}{\omega-v_i-U_i-U_{12}-\Delta_i}\nonumber\\
+\frac{U_{12}\braket{\braket{\hat n_{i\bar\sigma}n_{\bar i\sigma}\hat n_{\bar i\bar \sigma}\hat d_{i\sigma}:\hat d^{\dagger}_{i\sigma}}}_{\omega}}{\omega-v_i-U_i-U_{12}-\Delta_i} \;,
\label{eq_corr_three_body_2}
\end{gather}
and
\begin{gather}
\braket{\braket{\hat n_{i\bar\sigma}\hat n_{\bar i \bar \sigma }\hat d_{i\sigma}:\hat d^{\dagger}_{i\sigma}}}_{\omega}=\frac{\braket{\hat n_{i\bar\sigma}\hat n_{\bar i\bar \sigma}}}{\omega-v_i-U_i-U_{12}-\Delta_i}\nonumber\\
+\frac{U_{12}\braket{\braket{\hat n_{i\bar\sigma}n_{\bar i\sigma}\hat n_{\bar i\bar \sigma}\hat d_{i\sigma}:\hat d^{\dagger}_{i\sigma}}}_{\omega}}{\omega-v_i-U_i-U_{12}-\Delta_i} \;.
\label{eq_corr_three_body_3}
\end{gather}
Finally, the four-body GF reads
\begin{align}
  \braket{\braket{\hat n_{i\bar\sigma}\hat n_{\bar i\sigma}\hat n_{\bar i\bar \sigma}\hat d_{i\sigma}:\hat d^{\dagger}_{i\sigma}}}_{\omega}=\frac{\braket{\hat n_{i\bar\sigma}\hat n_{\bar i\bar\sigma}\hat n_{\bar i\sigma}}}{\omega-v_i-U_i-2U_{12}-\Delta_i}\;.
\label{eq_corr_four_body_1}
\end{align}
\cref{eq_corr_four_body_1} is the highest order GF in the EOM hierarchy. We
see that all the GFs depend not only on the parameters of the problem but also
on the various density correlators. In order to obtain expressions for the GFs,
we start by inserting the four-particle GF of \cref{eq_corr_four_body_1} into
Eqs.~(\ref{eq_corr_three_body_1})-(\ref{eq_corr_three_body_3}) for the
different three-particle GFs. Using partial fraction decompositions along the
way we obtain
\begin{gather}
  \braket{\braket{\hat n_{i\bar\sigma}\hat n_{\bar i \sigma }\hat d_{i\sigma}:\hat d^{\dagger}_{i\sigma}}}_{\omega} =
  \frac{\braket{\hat n_{i\bar\sigma}\hat n_{\bar i\sigma}} - \braket{\hat n_{i\bar\sigma}n_{\bar i\sigma}\hat n_{\bar i\bar \sigma}}}{\omega-v_i - U_i-U_{12}-\Delta_i}\nonumber\\
+\frac{\braket{\hat n_{i\bar\sigma}n_{\bar i\sigma}\hat n_{\bar i\bar \sigma}}}{\omega-v_i-U_i-U_{12}-\Delta_i} \;,
\label{eq_corr_three_body_1_2}
\end{gather}
\begin{gather}
\braket{\braket{\hat n_{\bar i\bar\sigma}\hat n_{\bar i \sigma }\hat d_{i\sigma}:\hat d^{\dagger}_{i\sigma}}}_{\omega}=\frac{\braket{\hat n_{\bar i\bar\sigma}\hat n_{\bar i\sigma}}-\braket{\hat n_{i\bar\sigma}n_{\bar i\sigma}\hat n_{\bar i\bar \sigma}}}{\omega-v_i-U_i-U_{12}-\Delta_i}\nonumber\\
+\frac{\braket{\hat n_{i\bar\sigma}n_{\bar i\sigma}\hat n_{\bar i\bar \sigma}}}{\omega-v_i-U_i-U_{12}-\Delta_i} \;,
\label{eq_corr_three_body_2_2}
\end{gather}
and
\begin{gather}
\braket{\braket{\hat n_{i\bar\sigma}\hat n_{\bar i \bar \sigma }\hat d_{i\sigma}:\hat d^{\dagger}_{i\sigma}}}_{\omega}=\frac{\braket{\hat n_{i\bar\sigma}\hat n_{\bar i\bar \sigma}}-\braket{\hat n_{i\bar\sigma}n_{\bar i\sigma}\hat n_{\bar i\bar \sigma}}}{\omega-v_i-U_i-U_{12}-\Delta_i}\nonumber\\
+\frac{\braket{\hat n_{i\bar\sigma}n_{\bar i\sigma}\hat n_{\bar i\bar \sigma}}}{\omega-v_i-U_i-U_{12}-\Delta_i} \;.
\label{eq_corr_three_body_3_2}
\end{gather}
Inserting Eqs.~(\ref{eq_corr_three_body_1_2})-(\ref{eq_corr_three_body_3_2})
into Eqs.~(\ref{eq_corr_two_body_1})-(\ref{eq_corr_two_body_3}) leads to the
two-particle GFs
\begin{gather}
  \braket{\braket{\hat n_{i\bar\sigma}\hat d_{i\sigma}:\hat d^{\dagger}_{i\sigma}}}_{\omega} \nonumber\\
  = \frac{\braket{\hat n_{ i\bar\sigma}} - \braket{\hat n_{ i\bar\sigma}\hat n_{ \bar i\sigma}} - \braket{\hat n_{ i\bar \sigma}\hat n_{ \bar i\bar\sigma}} +
    \braket{\hat n_{i\bar\sigma}n_{\bar i\sigma}\hat n_{\bar i\bar \sigma}}}{\omega-v_i-U_i-\Delta_i}\nonumber\\
  + \frac{\braket{\hat n_{ i\bar \sigma}\hat n_{ \bar i \sigma}} +
    \braket{\hat n_{ i\bar \sigma}\hat n_{ \bar i \bar \sigma}}-
 2 \braket{\hat n_{i\bar\sigma}n_{\bar i \sigma}\hat n_{\bar i\bar \sigma}}}{\omega-v_i-U_i - U_{12} -\Delta_i}\nonumber\\
+ \frac{\braket{\hat n_{i\bar\sigma}n_{\bar i\sigma}\hat n_{\bar i\bar \sigma}}}{\omega-v_i-U_i - 2 U_{12} - \Delta_i}\;,
  \label{eq_corr_two_body_1_2}
\end{gather}
\begin{gather}
  \braket{\braket{\hat n_{\bar i \sigma}\hat d_{i\sigma}:\hat d^{\dagger}_{i\sigma}}}_{\omega}\nonumber\\
  = \frac{\braket{\hat n_{\bar i \sigma}}- \braket{\hat n_{ i\bar\sigma}\hat n_{ \bar i\sigma}} - \braket{\hat n_{ \bar i\sigma}\hat n_{ \bar i\bar\sigma}} +
    \braket{\hat n_{i\bar\sigma}n_{\bar i\sigma}\hat n_{\bar i\bar \sigma}}}{\omega-v_i-U_{12}-\Delta_i}\nonumber\\
  +  \frac{\braket{\hat n_{i\bar\sigma}\hat n_{\bar i \sigma}} - \braket{\hat n_{i\bar\sigma}n_{\bar i\sigma}\hat n_{\bar i\bar \sigma}}}{\omega-v_i-U_i-U_{12}-\Delta_i}
  +  \frac{\braket{\hat n_{\bar i\sigma}\hat n_{\bar i \bar \sigma}} - \braket{\hat n_{i\bar\sigma}n_{\bar i\sigma}\hat n_{\bar i\bar \sigma}}}{\omega-v_i-2U_{12}-\Delta_i} 
  \nonumber\\
  +  \frac{\braket{\hat n_{i\bar\sigma}n_{\bar i\sigma}\hat n_{\bar i\bar \sigma}}}{\omega-v_i-U_i-2U_{12}-\Delta_i} \;,
  \label{eq_corr_two_body_2_2}
\end{gather}
and
\begin{gather}
  \braket{\braket{\hat n_{\bar i \bar \sigma}\hat d_{i\sigma}:\hat d^{\dagger}_{i\sigma}}}_{\omega}\nonumber\\
  = \frac{\braket{\hat n_{\bar i \bar \sigma}}- \braket{\hat n_{ i\bar\sigma}\hat n_{ \bar i\sigma}} - \braket{\hat n_{ \bar i\sigma}\hat n_{ \bar i\bar\sigma}} +
    \braket{\hat n_{i\bar\sigma}n_{\bar i\sigma}\hat n_{\bar i\bar \sigma}}}{\omega-v_i-U_{12}-\Delta_i}\nonumber\\
  +  \frac{\braket{\hat n_{i\bar\sigma}\hat n_{\bar i \sigma}} - \braket{\hat n_{i\bar\sigma}n_{\bar i\sigma}\hat n_{\bar i\bar \sigma}}}{\omega-v_i-U_i-U_{12}-\Delta_i}
  +  \frac{\braket{\hat n_{\bar i\sigma}\hat n_{\bar i \bar \sigma}} - \braket{\hat n_{i\bar\sigma}n_{\bar i\sigma}\hat n_{\bar i\bar \sigma}}}{\omega-v_i-2U_{12}-\Delta_i} 
  \nonumber\\
  +  \frac{\braket{\hat n_{i\bar\sigma}n_{\bar i\sigma}\hat n_{\bar i\bar \sigma}}}{\omega-v_i-U_i-2U_{12}-\Delta_i} \;.
  \label{eq_corr_two_body_3_2}
\end{gather}
Finally, in order to obtain the one-particle GF, we insert
Eqs.~(\ref{eq_corr_two_body_1_2})-(\ref{eq_corr_two_body_3_2}) into
Eq.~(\ref{eq_GF_one_body}) to arrive at Eq.~(\ref{eq_GF0}). In this very last
step we used some of the relations due to the spin symmetry of our problem
derived in Appendix \ref{app_correlators}. 

\section{Linear system for density correlators and analytic solution}
\label{app_correlators}

In the previous Appendix we have expressed the GFs in terms of density
correlators. The correlators themselves can be obtained from the GFs using
\cref{eq_exp_values}. Thus this latter relation allows to derive a linear system
for the correlators which can easily be solved numerically. Before deriving
this linear system, however, we show that as a consequence of the spin
symmetry in our problem, i.e., the fact that the on-site energies $v_i$ are
independent of $\sigma$, some of the density correlators coincide. This allows 
to reduce the size of the linear problem and for the case of our doube dot even
to find a fully analytic solution. 

We start by using \cref{eq_exp_values} on Eq.~(\ref{eq_corr_four_body_1})
with the embedding self energy in the wide-band limit
$\Delta_i = - i \frac{\gamma}{2}$. This leads to
\begin{equation}
  \braket{\hat n_{i\sigma}\hat n_{i\bar\sigma}\hat n_{\bar i\bar\sigma}\hat n_{\bar i\sigma}}
  = \phi(v_i + U_i + 2 U_{12}) 
  \braket{\hat n_{i\bar\sigma}\hat n_{\bar i\bar\sigma}\hat n_{\bar i\sigma}}\;,
  \label{4_part_correlator}
\end{equation}
where we used the function $\phi(p)$ defined in \cref{eq_int_digamma}. 
For given dot index $i$ the r.h.s. of \cref{4_part_correlator} is independent of
$\sigma$. Since the prefactor $\phi(v_i + U_i + 2 U_{12})$ on the l.h.s. of the
same equation is not only strictly positive at finite temperature but also
independent of $\sigma$, we conclude that the three-body correlator
$\braket{\hat n_{i\bar\sigma}\hat n_{\bar i\bar\sigma}\hat n_{\bar i\sigma}}$ is
independent of $\sigma$ as well. 

Using this result and a similar reasoning for
Eqs.~(\ref{eq_corr_three_body_1_2}) and (\ref{eq_corr_three_body_3_2}) we find
that all two-body correlators of the form
$\braket{\hat n_{1\sigma}\hat n_{2\sigma'}}$ are independent of both $\sigma$
and $\sigma'$. The spin symmetry of the two-body and the three-body
correlators can in an analogous way be used for
Eq.~(\ref{eq_corr_two_body_1_2}) to find that the spin densities
$\braket{\hat n_{i\sigma}}$ on dot $i$ are independent of $\sigma$ as expected.

If we use \cref{eq_exp_values} on Eqs.~(\ref{eq_corr_four_body_1}) -
(\ref{eq_corr_two_body_3_2}) as well as Eq.~(\ref{eq_GF0}) for all possible
values for the indices but do not make use of the spin symmetry we arrive at a
$15 \times 15$ linear system for the one-, two-, and three-body correlators.
Instead when using the spin symmetry relations derived above, the corresponding
linear system reduces to $7 \times 7$ thus leading to a significantly reduced
complexity. To be explicit, for the vector of density correlators
\begin{equation}
  {\mathbf{y}}= \left(
  \begin{array}{c}
    \braket{\hat n_1} \\
    \braket{\hat n_2} \\
    \braket{\hat n_{1\uparrow} \hat n_{1\downarrow}} \\
    \braket{\hat n_{2\uparrow} \hat n_{2\downarrow}} \\
    \braket{\hat n_{1\uparrow} \hat n_{2\downarrow}} \\
    \braket{\hat n_{1\uparrow} \hat n_{1\downarrow} \hat n_{2\uparrow}} \\
    \braket{\hat n_{2\uparrow} \hat n_{2\downarrow} \hat n_{1\uparrow}}
    \end{array}\right)\;,
\end{equation}
the linear system we need to solve reads
\begin{equation}
  {\mathbf{C y}} = {\mathbf{b}}\;,
  \label{linsys_correlators}
\end{equation}
where both the matrix $\mathbf{C}$ and the vector $\mathbf{b}$ are
explicitly given in terms of the model parameters as
\footnotesize

\begin{widetext}
	\begin{align}
{\mathbf{C}} = \begin{pmatrix}
	1+\ell_{11}-\ell_{12} & 2(\ell_{11}-\ell_{15}) & 0 & 2\ell_{15}-\ell_{11}-\ell_{16} & 2\zeta_1 & 0 & \nu_1\\
	2(\ell_{21}-\ell_{25}) & 1+\ell_{21}-\ell_{22} & 2\ell_{25}-\ell_{21}-\ell_{26} & 0 & 2\zeta_2 & \nu_2 & 0 \\
	-\ell_{12} & 0 & 1 & 0 & 2(\ell_{12}-\ell_{13}) & 0 & \ell_{13}-\ell_{12}-\ell_{14} \\
	0 & -\ell_{22} & 0 & 1 & 2(\ell_{22}-\ell_{23}) & \ell_{23}-\ell_{22}-\ell_{24} & 0 \\
	0 & -\ell_{15} & 0 & \ell_{15}-\ell_{16} & 1+\ell_{15}-\ell_{13} & 0 & \ell_{13}+\ell_{16}-\ell_{14}-\ell_{15} \\
	0 & 0 & 0 & 0 & -\ell_{13} & 1 & \ell_{13}-\ell_{14} \\
	0 & 0 & 0 & 0 & -\ell_{23} & \ell_{23}-\ell_{24} & 1
\end{pmatrix}\;,
\end{align}
\end{widetext}
\normalsize
where we have defined
\begin{subequations}
  \begin{align}
    \ell_{ij}= &\phi(p_{i,j}) \label{ell_ij}\;,\\
    \zeta_i= &\ell_{i2}+\ell_{i5}-\ell_{i1}-\ell_{i3}\;,\\
    \nu_i = &\ell_{i1}+2\ell_{i3}+\ell_{i6}-\ell_{i2}-\ell_{i4}-2\ell_{i5}\;,
    \end{align}
\end{subequations}
and
\begin{equation}
  {\mathbf{b}} =   \begin{pmatrix}
  	\phi(p_{1,1}) \\
  	\phi(p_{2,1})\\
  	0\\ 0\\ 0\\ 0\\ 0
  	\end{pmatrix}\;.
\end{equation}

While the linear system (\ref{linsys_correlators}) can easily be solved
numerically, one can also proceed to a fully analytical solution. To this end,
for a given dot index $i$ we first express the density correlators of
order higher than one in terms of the densities. This gives the
smaller linear system (note that for index $i$ the correlator
$\braket{\hat n_{i \sigma} \hat n_{i \bar \sigma} }$ does not appear)
\begin{align}
	\textbf{B}\begin{pmatrix}\braket{\hat n_{i\bar\sigma}\hat n_{\bar i\bar\sigma}\hat n_{\bar i\sigma}}\\
		\braket{\hat n_{i\sigma}\hat n_{i\bar\sigma}\hat n_{\bar i\sigma}} \\
		\braket{\hat n_{\bar i\sigma}\hat n_{\bar i\bar\sigma}}\\
		\braket{\hat n_{i\bar\sigma}\hat n_{\bar i\bar\sigma}}
	\end{pmatrix}
	=
	\begin{pmatrix}
		0\\
		0\\
		\phi(p_{\bar i,2})\braket{n_{\bar i\sigma}}\\
		\phi(p_{i,5})\braket{n_{\bar i\sigma}}
	\end{pmatrix}\;,
\end{align}
where the matrix ${\mathbf{B}}$ is given by 
\begin{widetext}
  \begin{align}
    \textbf{B}= \begin{pmatrix}
      1 & 0 & \ell_{i6}(\ell_{i4}-\ell_{i6}-1)^{-1} & 0 \\
      \ell_{i3}-\ell_{i4} & 1 & 0 & -\ell_{i3} \\
      0 & 0 & 1-\frac{\ell_{i6}(\ell_{\bar{i}2}+\ell_{\bar{i}4} -2\ell_{\bar{i}3})(\ell_{i4}-\ell_{i3})}{1-\ell_{i4}+\ell_{i6}} & 2(\ell_{\bar{i}2}-\ell_{\bar{i}3})+\ell_{i3}(2\ell_{\bar{i}3}-\ell_{\bar{i}2}-\ell_{\bar{i}4}) \\
      0 & 0 & \ell_{i5}-\ell_{i6}+\frac{\ell_{i6}(\ell_{i3}+\ell_{i6}-\ell_{i4}-\ell_{i5})}{1-\ell_{i4}+\ell_{i6}} & 1+\ell_{i5}-\ell_{i3}
    \end{pmatrix}\;.
  \end{align}
\end{widetext}
The solution of this smaller linear system reads

\begin{subequations}
  \begin{align}
    &\braket{\hat n_{\bar i\sigma}\hat n_{\bar i\bar\sigma}}=\tau_{i,1}\braket{\hat n_{\bar i\sigma}}\;,\\
    &\braket{\hat n_{i\bar\sigma}\hat n_{\bar i\bar\sigma}}=\tau_{i,2}\braket{\hat n_{\bar i\sigma}}\;,\\
    &	\braket{\hat n_{i\bar\sigma}\hat n_{\bar i\bar\sigma}\hat n_{\bar i\sigma}}=\tau_{i,3}\braket{\hat n_{\bar i\sigma}}\;,\\
    &	\braket{\hat n_{i\sigma}\hat n_{i\bar\sigma}\hat n_{\bar i\sigma}} =\tau_{i,4}\braket{\hat n_{\bar i\sigma}}\;,
  \end{align}
  \label{eq_solution_two_body_2}
\end{subequations}
where  
\begin{widetext}
	\begin{subequations}
		\begin{align}
			\tau_{i,1} = &\frac{1}{\tau_{i,0}}\left[\ell_{\bar{i}2}\left(\ell_{i3}-1\right)\left(\ell_{i5}-1\right)+2\left(1-\ell_{i3}\right)\ell_{\bar{i}3}\ell_{i5}+\ell_{i3}\ell_{\bar{i}4}\ell_{i5}\right] \;,\\
			\tau_{i,2} = &\frac{1}{\tau_{i,0}}\left[\ell_{\bar{i}2}\left(\ell_{i6}+\ell_{i5}\left(\frac{\ell_{i6}}{1-\ell_{i4}+\ell_{i6}}-1\right)-\frac{\ell_{i6}}{1-\ell_{i4}+\ell_{i6}}\left(\ell_{i3}-\ell_{i4}+\ell_{i6}\right)\right)\right]\nonumber\\
			&+\frac{1}{\tau_{i,0}}\left[1-\frac{\ell_{i6}}{1-\ell_{i4}+\ell_{i6}}\left(\ell_{\bar{i}2}+\ell_{\bar{i}4}-2\ell_{\bar{i}3}\right)\left(\ell_{i4}-\ell_{i3}\right)\right]\ell_{i5}\;, \\
			\tau_{i,3} = &\frac{\ell_{i6}}{1-\ell_{i4}+\ell_{i6}}\tau_{i,1} \;,\\
			\tau_{i,4} = &\left(\ell_{i4}-\ell_{i3}\right)\frac{\ell_{i6}}{1-\ell_{i4}+\ell_{i6}}\tau_{i,1}+\ell_{i3}\tau_{i,2} \;,\\
			\tau_{i,0} = &\left(\ell_{\bar{i}2}(\ell_{i3}-2)+2(1-\ell_{i3})\ell_{\bar{i}3}+\ell_{i3}\ell_{\bar{i}4}\right)\left(\ell_{i5}-\ell_{i6}+\frac{\ell_{i6}}{1-\ell_{i4}+\ell_{i6}}\left(\ell_{i3}-\ell_{i4}-\ell_{i5}+\ell_{i6}\right)\right)\nonumber\\
			&+\left[1-\frac{\ell_{i6}}{1-\ell_{i4}+\ell_{i6}}\left(\ell_{\bar{i}2}+\ell_{\bar{i}4}-2\ell_{\bar{i}3}\right)\left(\ell_{i4}-\ell_{i3}\right)\right](1+\ell_{i5}-\ell_{i3})\;,
		\end{align}
		\label{eq_iota_coeff}
	\end{subequations}
\end{widetext}
where the $\ell_{ij}$ are defined in \cref{ell_ij}.
Since in this way, all the density correlators (and thus all the residues
of the single-particle GF, see \cref{eq_numerators_one_body}) are expressed in
terms of the densities themselves, the final step to obtain the densities
requires the solution of the $2 \times 2$ linear system
\begin{align}
\begin{pmatrix}
\eta_{11} &\eta_{12} \\
\eta_{21} & \eta_{22} 
\end{pmatrix}
\begin{pmatrix}
\braket{\hat n_{i\sigma}}  \\
\braket{\hat n_{\bar i\sigma}}  
\end{pmatrix}=
\begin{pmatrix}
\phi(p_{1,1})  \\
\phi(p_{2,1})
\end{pmatrix}\;,
\end{align}
where the parameters $\eta_{i,j}$ are given in \cref{eq_eta_params}. The
solution of the linear system finally gives the expression for the densities
\cref{eq_local_occupations}.

\acknowledgments
We acknowledge useful discussions with Gianluca Stefanucci. 
We acknowledge financial support through Grant PID2020-112811GB-I00 funded by
MCIN/AEI/10.13039/501100011033  as well as by grant IT1453-22 “Grupos
Consolidados UPV/EHU  del Gobierno  Vasco”. We acknowledge the technical support provided by SGIker (Scientific Computing Services UPV/EHU).

\bibliography{biblio}

\end{document}